\documentclass{article}
\usepackage{spconf,amsmath,graphicx,amssymb}
\usepackage{spconf,amsmath,graphicx,amssymb,mathrsfs,balance,diagbox,appendix,booktabs}
\usepackage{bbding}
\usepackage{pifont}
\usepackage{wasysym}
\usepackage{amssymb}
\usepackage{balance}
\usepackage{setspace}
\usepackage[colorlinks,linkcolor=black]{hyperref}
\usepackage{underscore}
\usepackage[utf8]{inputenc}

\title{Uformer: A Unet based dilated complex \& real dual-path conformer network for simultaneous speech enhancement and dereverberation}
\name{Yihui Fu$^{1}$, Yun Liu$^{2}$, Jingdong Li$^{2}$, Dawei Luo$^{2}$, Shubo Lv$^{1}$, Yukai Jv$^{1}$, Lei Xie$^{1,*}\thanks{* Corresponding author.}$}
\address{Audio, Speech and Language Processing Group (ASLP@NPU), \\ Northwestern Polytechnical University, Xi'an, China\\
AI Interaction Division, Sogou Inc., Beijing, China}

\begin{document}
\ninept
\maketitle
%$\thanks{* This work is done during the internship in Sogou Inc.}, 
%
% \begin{abstract}
% In this paper, we propose Uformer, a Unet based dilated complex \& real dual-path conformer network in both magnitude and complex domain for simultaneous speech enhancement and dereverberation, as noise and reverberation are regarded as two major factors that degrade the speech quality. We exploit time attention (TA) and dilated convolution (DC) to leverage local and global contextual information as well as frequency attention (FA) to model dimentional information within the proposed dilated complex \& real dual-path conformer module to improve the speech enhancement and dereverberation ability. Furthermore, hybrid encoder and decoder is adopted, which models the complex spectrum and magnitude simultaneously and promote the information exchange between two domains. Encoder-decoder attention is applied to better model the information interaction between encoder and decoder. Our experimental results indicate that the proposed method reaches 3.6032 DNSMOS on the blind test set of Interspeech 2021 DNS Challenge, which outperforms all SOTA time domain and complex domain speech front-end models on both objective and subjective performance.
% \end{abstract}
\begin{abstract}
\vspace{-0.1cm}
% Complex spectrum and magnitude are recognized as two major features of speech enhancement and dereverberation. Traditionally, the focus has always been treating these two features separately, ignoring their underlying relationship. In this paper, we propose Uformer, a Unet based dilated complex \& real dual-path conformer network in both complex and magnitude domain for simultaneous speech enhancement and dereverberation. We exploit time attention (TA) and dilated convolution (DC) to leverage local and global contextual information and frequency attention (FA) to model dimensional information. These three sub-modules contained in the proposed dilated complex \& real dual-path conformer module effectively improve the speech enhancement and dereverberation performance. Furthermore, hybrid encoder and decoder are adopted to simultaneously model the complex spectrum and magnitude and promote the information interaction between two domains. Encoder decoder attention is also applied to enhance the interaction between encoder and decoder. Our experimental results outperform all SOTA time domain and complex domain speech front-end models on both objective and subjective performance. Specifically, Uformer reaches 3.6032 DNSMOS on the blind test set of Interspeech 2021 DNS Challenge, which outperforms all SOTA time domain and complex domain speech front-end models on both objective and subjective performance. We also carry out ablation experiments to tease apart all proposed sub-modules that are most important.
Complex spectrum and magnitude are considered as two major features of speech enhancement and dereverberation. Traditional approaches always treat these two features separately, ignoring their underlying relationship. In this paper, we propose \textit{Uformer}, a Unet based dilated complex \& real dual-path conformer network in both complex and magnitude domain for simultaneous speech enhancement and dereverberation. We exploit time attention (TA) and dilated convolution (DC) to leverage local and global contextual information and frequency attention (FA) to model dimensional information. These three sub-modules contained in the proposed dilated complex \& real dual-path conformer module effectively improve the speech enhancement and dereverberation performance. Furthermore, hybrid encoder and decoder are adopted to simultaneously model the complex spectrum and magnitude and promote the information interaction between two domains. Encoder decoder attention is also applied to enhance the interaction between encoder and decoder. Our experimental results outperform all SOTA time and complex domain models objectively and subjectively. Specifically, Uformer reaches 3.6032 DNSMOS on the blind test set of Interspeech 2021 DNS Challenge, which outperforms all top-performed models. We also carry out ablation experiments to tease apart all proposed sub-modules that are most important.

\end{abstract}
\vspace{-0.15cm}
\begin{keywords}
speech enhancement and dereverberation, Uformer, dilated complex dual-path conformer, hybrid encoder and decoder, encoder decoder attention
\end{keywords}
\vspace{-0.4cm}
\section{Introduction}
\vspace{-0.3cm}

Owing to the success of deep learning, deep neural network (DNN) based speech enhancement and dereverberation have progressed dramatically.
For a long time, DNN-based speech enhancement algorithms attempt to only enhance the noisy magnitude using ideal ratio mask (IRM) and keep the noisy phase when reconstructing speech waveform~\cite{wang2014training}. This operation can be attributed to the unclear structure of phase, which is considered difficult to estimate~\cite{Wang1982The}. 
Then, literature has emerged that offers contradictory findings about the importance of phase from perception aspect~\cite{paliwal2011importance}. 
Subsequently, complex ratio mask (CRM) was proposed by Williamson \emph{et al.}~\cite{williamson2017time}, reconstructing speech perfectly by enhancing both real and imaginary components of the noisy speech simultaneously. 
Later, Tan \emph{et al.} proposed a convolution recurrent network (CRN) with one encoder and two decoders for complex spectral mapping (CSM) to estimate the real and imaginary components of mixture speech simultaneously~\cite{tan2019learning}. It is worth considering that CRM and CSM possess the full information (magnitude and phase) of speech signal so that they can achieve the best oracle speech enhancement performance in theory. 
Then, the study in~\cite{hu2020dccrn} updated the CRN network by introducing complex convolution and LSTM, resulting in the deep complex convolution recurrent network (DCCRN), which ranked first in MOS evaluation in the Interspeech 2020 DNS challenge real-time-track. 
The study in~\cite{lv2021dccrn+} further optimized DCCRN by proposing DCCRN+, which enables the model with subband processing ability by learnable neural network filters. It formulated the decoder as a multi-task learning framework with an auxiliary task of a priori SNR estimation. 
Besides speech enhancement, dereverberation is also a challenging task since it is hard to pinpoint the direct path signal and differentiate it from its copies, especially when reverberation is strong and non-stationary noise is also present. Some previous works have already been done on simultaneous multi-channel enhancement and dereverberation~\cite{kagami2018joint,gode2021joint,pfeifenberger2021blind}. However, it is still a relatively hard task for the single-channel scenario due to the lack of spatial information. In this scenario, the feature they can only use is single-channel magnitude or complex spectrum. Although some researches exploit magnitude and complex spectrum as two input features, no parameter interaction between them are conducted. %So we break the trend of recent methods that are only built on these two separate features. Novelly, we apply a hybrid encoder and decoder to model complex spectrum and magnitude simultaneously.
% On the other hand, conformer~\cite{gulati2020conformer} is an outstanding convolution-augmented transformer model for speech recognition with strong contextual modeling ability. The general conformer model consists of encoder and decoder networks. %In our work, we only use the encoder part since the input mixtures and output enhanced sequences have the same length in speech front-end processing.
% The original conformer model consists of three main parts, named feed forward module, self attention module and convolution module. The convolution module captures the local interaction of the features, where the self attention module capture the global feature interactions. Recently, researchers have introduced self attention based model like Transformer to front-end task. Dang et al. and Wang et al. propose DPT-FSNET~\cite{dang2021dpt} and TSTNN~\cite{wang2021tstnn} respectively, which adopted dual-path transformer to model local and global information of sequences. TeCANet~\cite{wang2021tecanet} applied transformer to the frames within a context window thus estimating correlations between frames. However, the strong temporal modeling ability of Conformer is still to be explored.

Updated from transformer~\cite{vaswani2017attention}, conformer~\cite{gulati2020conformer} is an outstanding convolution-augmented model for speech recognition with strong contextual modeling ability. The original conformer model presents two important modules: self-attention module and convolution module to model the global and local information, respectively. 
These modules have also proven their effectiveness, especially for temporal modeling ability from speech separation task~\cite{ chen2021continuous}.
Other transformer-based models, e.g., DPT-FSNET~\cite{dang2021dpt} and TSTNN~\cite{wang2021tstnn}, adopted dual-path transformer, also showed stronger interpretability and temporal-dimensional modeling ability than single-path method.
TeCANet~\cite{wang2021tecanet} applied transformer to the frames within a context window thus estimating correlations between frames. To further achieve stronger contextual modeling ability, the combination of conformer and dual-path method is an instinctive idea.

\begin{figure}[t]
		\centering
	    \includegraphics[width=0.5 \textwidth]{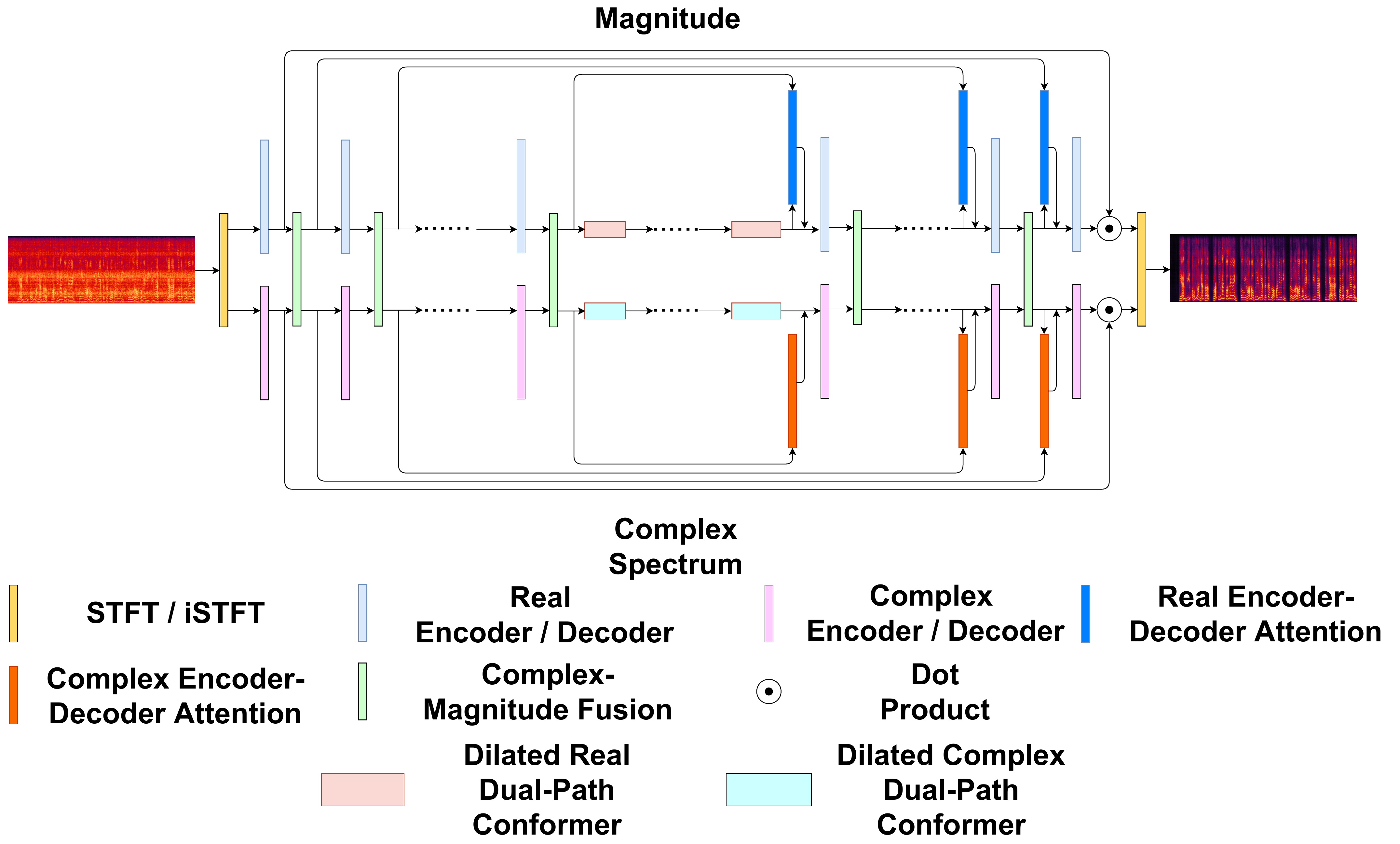}
		%\centerline{\includegraphics[width=10.0cm]{figures/topo}}  %,trim=0 40 0 0
    %\setlength{\abovecaptionskip}{-1.5cm}
    \vspace{-0.7cm}
	\caption{The overall architecture of Uformer.}
	\label{overall}
\end{figure}

Inspired by the considerations mentioned above, we propose \emph{Uformer}, a Unet based dilated complex \& real dual-path conformer network for simultaneous speech enhancement and dereverberation. In magnitude branch, we seek to construct the filtering system which only applies to the magnitude domain. Specifically in the magnitude branch, we seek to construct a filtering system which only applies to the magnitude domain. In this branch, most of the noise is expected to be effectively suppressed. By contrast, the complex domain branch is established as a decorating system to compensate for the possible loss of spectral details and phase mismatch. Two branches work collaboratively to facilitate the overall spectrum recovery. The overall architecture of Uformer is shown in Figure~\ref{overall}. Our contribution in this work is three-fold: 
\vspace{-0.1cm}
\begin{itemize}
    \setlength{\topmargin}{0pt}
    \setlength{\itemsep}{0em}
    \setlength{\parskip}{0pt}
    \setlength{\parsep}{0pt}
    \item We apply \emph{dilated complex \& real dual-path conformer} on the bottle-neck feature between encoder and decoder. Within the proposed module, \emph{Time attention (TA)} with a context window is applied to model the local time dependency while \emph{dilated convolution (DC)} is used to model global time dependency. \emph{Frequency attention (FA)} is used to model subband information.

    \item We apply a \emph{hybrid encoder and decoder} to model complex spectrum and magnitude simultaneously. The rationale is that there exists close relation between magnitude and phase in complex spectrum. Superb magnitude estimation can profit better recovery for phase and vice versa.

    \item We utilize \emph{encoder decoder attention} to estimate attention mask, instead of skip connection, to reveal the relevance between the corresponding hybrid encoder and decoder layers.
\end{itemize}
\vspace{-0.1cm}
Our experimental results outperform all SOTA time domain and complex domain speech front-end models on both objective and subjective performance. Specifically, Uformer reaches 3.6032 DNSMOS on the blind test set of Interspeech 2021 DNS challenge, which surpass SDD-Net~\cite{li2021simultaneous}, the top-performed model in the challenge. We also carry out ablation study to tease apart all proposed sub-modules that are most important.

\vspace{-0.3cm}

\begin{figure}[t]
		\centering
	    \includegraphics[width=0.45 \textwidth]{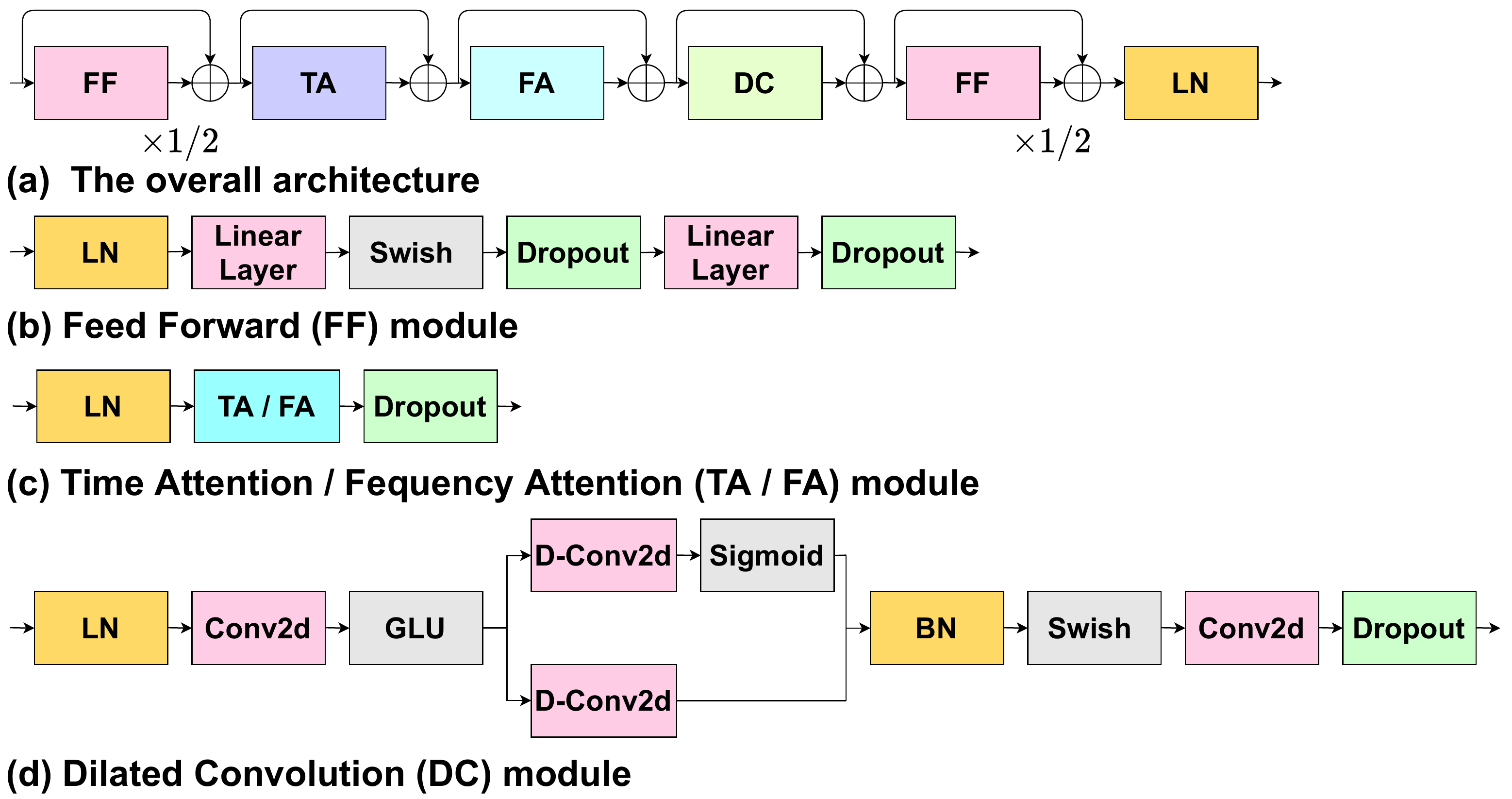}
		%\centerline{\includegraphics[width=10.0cm]{figures/topo}}  %,trim=0 40 0 0
    %\setlength{\abovecaptionskip}{-1.5cm}
    \vspace{-0.3cm}
	\caption{The architecture of dilated complex dual-path conformer.}
	\label{conformer}
	\vspace{-0.3cm}
\end{figure}

\vspace{-0.15cm}
\section{Proposed method}
\vspace{-0.3cm}
\subsection{Problem Formulation}
\vspace{-0.2cm}
In the time domain, let $\mathbf{s}(t)$, $\mathbf{h}(t)$, $\mathbf{n}(t)$ denote the anechoic speech, room impulsive reponse (RIR) and noise, respectively. The observed signal of microphone $\mathbf{o}(t)$ can be denoted as:
\vspace{-0.2cm}
\begin{equation}
\mathbf{o}(t) = \mathbf{s}(t)*\mathbf{h}(t)+\mathbf{n}(t) = \mathbf{s_{e}}(t) + \mathbf{s_{l}}(t) + \mathbf{n}(t),
\vspace{-0.2cm}
\end{equation}
where $*$ denotes convolution operation, $\mathbf{s_{e}}(t)$ refers to direct sound plus early reflections, and $\mathbf{s_{l}}(t)$ denotes late reverberation. The corresponding frequency domain signal model is shown as:
\vspace{-0.2cm}
\begin{equation}
\mathbf{O}(t,f) = \mathbf{S_{e}}(t,f)+\mathbf{S_{l}}(t,f)+\mathbf{N}(t,f),
\vspace{-0.2cm}
\end{equation}
where $\mathbf{O}(t,f)$, $\mathbf{S_{e}}(t,f)$, $\mathbf{S_{l}}(t,f)$ and $\mathbf{N}(t,f)$ are the corresponding variable with frame index $t$ and frequency index $f$ in frequency domain. Our target is to estimate the $\mathbf{s_{e}}$ and $\mathbf{S_{e}}$ in this work.
\vspace{-0.43cm}
\subsection{Complex Self Attention}
\vspace{-0.2cm}
In the original self attention, the input $\mathbf{X }$ is mapped with different learnable linear transformation $\mathbf{W}$ to get queries ($\mathbf{Q}$), keys ($\mathbf{K}$) and values ($\mathbf{V}$) respectively:
\vspace{-0.2cm}
\begin{equation}
\label{qkv}
\mathbf{Q} = \mathbf{X}\mathbf{W}_{Q}, \mathbf{K} = \mathbf{X}\mathbf{W}_{K}, \mathbf{V} = \mathbf{X}\mathbf{W}_{V}.
\vspace{-0.25cm}
\end{equation}
Then, the dot product results of queries with keys are computed, followed by division of a constant value $k$, representing the projection dimension of $\mathbf{W}$. After applying the softmax function to generate the weights, the weighted result is obtained by:
\vspace{-0.25cm}
\begin{equation}
\label{weight}
\text{Attention}(\mathbf{Q},\mathbf{K},\mathbf{V}) = \text{softmax}(\frac{\mathbf{Q}^\mathsf{T} \mathbf{K}}{\sqrt{k}}) \mathbf{V}.
\vspace{-0.25cm}
\end{equation}
% Inspired by~\cite{yang2020complex}, given the complex input $\mathbf{X}$, the complex valued $\mathbf{Q}$ is calculated by:
Inspired by~\cite{yang2020complex}, given the complex input $\mathbf{X}$, the complex valued $\mathbf{Q}$ is calculated by:
\vspace{-0.2cm}
\begin{equation}
\begin{aligned}
\mathbf{Q}^{\Re} &= \mathbf{X}^{\Re}\mathbf{W}_{Q}^{\Re} - \mathbf{X}^{\Im}\mathbf{W}_{Q}^{\Im}, \\
\mathbf{Q}^{\Im} &= \mathbf{X}^{\Re}\mathbf{W}_{Q}^{\Im} + \mathbf{X}^{\Im}\mathbf{W}_{Q}^{\Re},
\vspace{-0.2cm}
\end{aligned}
\end{equation}
where $\Re$ and $\Im$ indicate the real and imaginary parts, respectively. $\mathbf{K}$ and $\mathbf{V}$ are calculated in the same way. Thus, the complex self attention is calculated by:
\vspace{-0.2cm}
\begin{equation}
\begin{aligned}
&\text{ComplexAttention}(\mathbf{Q},\mathbf{K},\mathbf{V}) = \\
(&\text{Attention}(\mathbf{Q}^{\Re},\mathbf{K}^{\Re},\mathbf{V}^{\Re}) - \text{Attention}(\mathbf{Q}^{\Re},\mathbf{K}^{\Im},\mathbf{V}^{\Im}) - \\
&\text{Attention}(\mathbf{Q}^{\Im},\mathbf{K}^{\Re},\mathbf{V}^{\Im})- \text{Attention}(\mathbf{Q}^{\Im},\mathbf{K}^{\Im},\mathbf{V}^{\Re})) + \\
i(&\text{Attention}(\mathbf{Q}^{\Re},\mathbf{K}^{\Re},\mathbf{V}^{\Im}) + \text{Attention}(\mathbf{Q}^{\Re},\mathbf{K}^{\Im},\mathbf{V}^{\Re}) +\\
&\text{Attention}(\mathbf{Q}^{\Im},\mathbf{K}^{\Re},\mathbf{V}^{\Re})- \text{Attention}(\mathbf{Q}^{\Im},\mathbf{K}^{\Im},\mathbf{V}^{\Im})).
\end{aligned}
\end{equation}
\vspace{-0.8cm}
\subsection{Dilated Complex Dual-path Conformer}
\vspace{-0.2cm}
The overall architecture of dilated complex dual-path conformer is shown Figure~\ref{conformer} (a). The module consists of time attention (TA), frequency attention (FA), dilated convolution (DC) and two feed forward (FF) layers. As for dilated real dual-path conformer, all modules have the same real-valued architectures with dilated complex dual-path conformer. 

As shown in Figure~\ref{conformer} (b), FF consists of two linear layers with a swish activation in between. Inspired by~\cite{gulati2020conformer}, we employ half-step residual weights in our FF.

In the indoor acoustic scenario, sudden noises like clicking, coughing, etc., are common noise types without very long contextual dependency. On the other hand, the received signal is a collection of many delayed and attenuated copies of the original speech signal, resulting in strong feature correlations in a restricted range of time steps of the input sequence. To better model the local temporal relevance information mentioned above, inspired by~\cite{wang2021tecanet}, we use TA, which utilizes the frames within a context window, instead of the whole sentence, to calculate the local temporal features. For the input feature of TA: $\mathbf{X}_{TA}(t, f) \in \mathbb{C}^{C}$ in Figure~\ref{conformer} (c), a $c$-frame expansion $\overline{\mathbf{X}_{TA}}(t, f) \in \mathbb{C}^{c \times C}$ is generated as contextual feature, where $C$ denotes channel numbers of bottle-neck feature. The $\mathbf{Q}_{TA}(t, f) \in \mathbb{C}^{d_{TA} \times c}$, $\mathbf{K}_{TA}(t, f) \in \mathbb{C}^{d_{TA} \times 1}$ and $\mathbf{V}_{TA}(t, f) \in \mathbb{C}^{c \times d_{TA}}$ are obtained from $\overline{\mathbf{X}_{TA}}$, $\mathbf{X}_{TA}$ and $\overline{\mathbf{X}_{TA}}$ respectively by Eq.~(\ref{qkv}). The weight distribution on the context frames is computed based on the similarities between $\mathbf{Q}_{TA}(t, f)$ and 
$\mathbf{K}_{TA}(t, f)$ by the scaled dot product:
\vspace{-0.2cm}
\begin{equation}
\mathbf{A}_{TA}(t,f) = \text{softmax}(\frac{\mathbf{Q}_{TA}^\mathsf{T} \mathbf{K}_{TA}}{\sqrt{d_{TA}}}).
\vspace{-0.2cm}
\end{equation}
Multiplying the weight $\mathbf{A}_{TA}(t,f) \in \mathbb{C}^{c \times 1}$ with $\mathbf{V}(i)$, we can get the weighted feature with local temporal relevance information:
\vspace{-0.2cm}
\begin{equation}
\mathbf{Y}_{TA}(t,f) = \mathbf{A}_{TA}(t,f) \odot \mathbf{V}_{TA}(t,f),
\vspace{-0.2cm}
\end{equation}
where $\odot$ denotes the dot product. A fully connected layer is followed to project $\mathbf{Y}_{TA}$ to the same dimension with $\mathbf{X}_{TA}$.

The lower frequency bands tend to contain high energies while the higher frequency bands tend to contain low energies. Therefore, we also pay different attention to different frequency subbands. In our proposed frequency attention (FA), for the input feature  $\mathbf{X}_{FA}(t) \in \mathbb{C}^{F \times C}$ in Figure~\ref{conformer} (c), projected features $\mathbf{Q}_{FA}(t) \in \mathbb{C}^{d_{FA} \times F}$, $\mathbf{K}_{FA}(t) \in \mathbb{C}^{d_{FA} \times F}$ and $\mathbf{V}_{FA}(t) \in \mathbb{C}^{F\times d_{FA}}$ are obtained by Eq.~(\ref{qkv}). After generating the weight distribution on different subbands $\mathbf{A}(t)_{FA} \in \mathbb{C}^{F \times F}$, the FA result $\mathbf{Y}(t)_{FA}\in \mathbb{C}^{F \times d_{FA}}$ is calculated by Eq.~(\ref{weight}). A fully connected layer is followed to project $\mathbf{Y}_{TA}$ to the same dimension with $\mathbf{X}_{FA}$. The detailed architecture of TA and FA are shown in Figure~\ref{conformer} (c).

TasNet~\cite{luo2019conv} has shown superior performance in speech separation, which uses stacked temporal convolution network (TCN) to better capture long range sequence dependencies. Original TCN first projects the input to a higher channel space with Conv1d. Then a dilated depthwise convolution (D-Conv) is applied to get a larger receptive field. The output Conv1d projects the number of channel the same as the input. Residual connection is applied to enforce the network to focus on the missing detail and mitigate gradient disappearance. Our improvements on TCN in dilated dual-path conformer are shown in Figure~\ref{conformer} (d). Gated D-Conv2d is applied with opposite dilation in two D-Conv2ds. The dilation of lower D-Conv2d is $1,2,...,2^{N-1}$ while the dilation of upper gated D-Conv2d is $2^{N-1}, 2^{N-2},..., 1$, where $N$ denotes the cascaded layer number of dilated dual-path conformer.

\vspace{-0.4cm}
\subsection{Hybrid Encoder and Decoder}
\vspace{-0.2cm}
In our proposed hybrid encoder and decoder, both encoder and decoder model the complex spectrum and magnitude at the same time. Let $\mathbf{C}_{i}$ and $\mathbf{M}_{i}$ denotes the complex spectrum and magnitude output of encoder/decoder layer $i$ respectively. To promote the information exchange between complex spectrum and magnitude, the complex-magnitude fusion results $\mathbf{\hat{C}}_i$ and $\mathbf{\hat{M}}_i$ are calculated by:
\vspace{-0.2cm}
\begin{equation}
\begin{aligned}
\mathbf{\hat{C}}_i^{\Re } &= \mathbf{C}_i^{\Re}+\sigma(\mathbf{M}_{i}), \\
\mathbf{\hat{C}}_i^{\Im } &= \mathbf{C}_i^{\Im}+\sigma(\mathbf{M}_{i}), \\
\vspace{-0.2cm}
\mathbf{\hat{M}}_i &= \mathbf{M}_{i}+\sigma(\sqrt{\mathbf{C}_i^{\Re^{2}}+\mathbf{C}_i^{\Im^{2}}}).
\end{aligned}
\end{equation}
% \begin{equation}

% \end{equation}
% \begin{equation}

% \end{equation}
\vspace{-0.7cm}
\subsection{Encoder Decoder Attention}
\vspace{-0.2cm}
Let $\mathbf{E}_{i}$ denotes the output of hybrid encoder layer $i$ while $\mathbf{D}_{i}$ denotes the output of dilated conformer layer or hybrid decoder layer $i$. Two Conv2ds are first applied to $\mathbf{E}_{i}$ and $\mathbf{D}_{i}$ to map them to high dimensional features $\mathbf{G}_{i}$ by:
\vspace{-0.2cm}
\begin{equation}
\mathbf{G}_{i} = \sigma(\mathbf{W}^{E}_{i} \ast  \mathbf{E}_{i} + \mathbf{W}^{D}_{i} \ast  \mathbf{D}_{i}),
\vspace{-0.2cm}
\end{equation}
where $\mathbf{W}^{E}_{i}$ and $\mathbf{W}^{D}_{i}$ denote the convolution kernel first two Conv2ds of layer $i$, respectively. After applying a third Conv2d to $\mathbf{G}_{i}$, the final output is calculated by multiplying the sigmoid attention mask with $\mathbf{D}_{i}$:
\vspace{-0.2cm}
\begin{equation}
\mathbf{\hat{D}}_{i} = \sigma(\mathbf{W}^{A}_{i} \ast  \mathbf{G}_{i}) \odot \mathbf{D}_{i},
\vspace{-0.2cm}
\end{equation}
where $\mathbf{W}^{A}_{i}$ denotes the convolution kernel the third Conv2d of layer $i$.
Finally, we concatenate $\mathbf{D}_{i}$ and $\mathbf{\hat{D}}_{i}$ along channel axis as the input of the next decoder layer.
\vspace{-0.45cm}
\subsection{Loss Function}
\vspace{-0.25cm}
After generating the estimated CRM $\mathbf{H}_{C}$ and IRM $\mathbf{H}_{R}$ from the last decoder layer, we multiply the noisy complex spectrum and magnitude with them respectively to get enhanced and dereverbed complex spectrum and magnitude:
\vspace{-0.2cm}
\begin{equation}
\mathbf{M}^{\text{mag}}_{C} =\tanh(\sqrt{\mathbf{H}^{\Re 2}_{C}+\mathbf{H}^{\Im 2}_{C}}), 
\vspace{-0.5cm}
\end{equation}

\begin{equation}
\mathbf{M}^{\text{pha}}_{C} =\text{arctan2}(\mathbf{H}^{\Im}_{{C}}, \mathbf{H}^{\Re}_{{C}}), 
\end{equation}
\begin{equation}
\mathbf{M}^{\text{mag}}_{R} = \sigma(\mathbf{H}_{R}),  
\end{equation}
\begin{equation}
\mathbf{Y}^{\text{mag}}_{C} = \mathbf{X}^{\text{mag}}  \odot \mathbf{M}^{\text{mag}}_{C},  
\end{equation}
\begin{equation}
\mathbf{Y}^{\text{pha}}_{C} = \mathbf{X}^{\text{pha}}  + \mathbf{M}^{\text{pha}}_{C}, 
\end{equation}
\begin{equation}
\mathbf{Y}^{\text{mag}}_{R} = \mathbf{X}^{\text{mag}}  \odot \mathbf{M}^{\text{mag}}_{{R}}, 
\end{equation}
\begin{equation}
\overline{\mathbf{Y}^{\text{mag}}_{C}} = (\mathbf{Y}^{\text{mag}}_{C} +\mathbf{Y}^{\text{mag}}_{R}) / 2, 
\label{outputfusion}
\end{equation}
\begin{equation}
\mathbf{Y}^{\Re} =  \overline{\mathbf{Y}^{\text{mag}}_{C}} \odot \cos(\mathbf{Y}^{\text{pha}}_{C}), 
\end{equation}
\begin{equation}
\mathbf{Y}^{\Im} =  \overline{\mathbf{Y}^{\text{mag}}_{C}} \odot \sin(\mathbf{Y}^{\text{pha}}_{C}), 
\end{equation}
\begin{equation}
\mathbf{y} = \text{iSTFT}(\mathbf{Y}^{\Re}, \mathbf{Y}^{\Im}). 
\vspace{-0.2cm}
\end{equation}

We use hybrid time and frequency domain loss as the target:
\vspace{-0.2cm}
\begin{equation}
\mathcal{L} = \alpha \mathcal{L}_{\textbf{SI-SNR}} + \beta \mathcal{L}_{\textbf{L1}}^{\textbf{T}} + \gamma \mathcal{L}_{\textbf{L2}}^{\textbf{C}} + \zeta \mathcal{L}_{\textbf{L2}}^{\textbf{M}},
\vspace{-0.2cm}
\end{equation}
where $\mathcal{L}_{\textbf{SI-SNR}}$, $\mathcal{L}_{\textbf{L1}}^{\textbf{T}}$, $\mathcal{L}_{\textbf{L2}}^{\textbf{C}}$ and $\mathcal{L}_{\textbf{L2}}^{\textbf{M}}$ denote SI-SNR~\cite{le2019sdr} loss in time domain, L1 loss in time domain, complex spectrum L2 loss and magnitude L2 loss, respectively. $\alpha$, $\beta$, $\gamma$ and $\zeta$ denote the weight of four losses. Four losses are calculated by:
\vspace{-0.4cm}
\begin{equation}
\begin{aligned}
\mathcal{L}_{\textbf{SI-SNR}} &= -20\log_{10}\frac{\Vert \xi  \cdot \mathbf{\hat{y}} \Vert}{\Vert \mathbf{y} -\xi  \cdot \mathbf{\hat{y}} \Vert}, \\
\xi  &= \mathbf{y}^\text{T}\mathbf{\hat{y}}/\mathbf{\hat{y}}^\text{T}\mathbf{\hat{y}},
\end{aligned}
\end{equation}
\vspace{-0.2cm}
\begin{equation}
\mathcal{L}_{\textbf{L1}}^{\textbf{T}} = \sum_{t} \left | (\mathbf{y}(t) - \mathbf{\hat{y}}(t))\right |,
\end{equation}
\vspace{-0.3cm}
\begin{equation}
\begin{aligned}
\mathcal{L}_{\textbf{L2}}^{\textbf{C}} = (\sum_{t,f}& \left | (\mathbf{Y}^{\Re}(t,f) - \mathbf{\hat{Y}}^{\Re}(t,f))\right | ^2 +\\
& \left | (\mathbf{Y}^{\Im}(t,f) - \mathbf{\hat{Y}}^{\Im}(t,f))\right | ^2) / F,
\end{aligned}
\end{equation}
\vspace{-0.1cm}
\begin{equation}
\mathcal{L}_{\textbf{L2}}^{\textbf{M}} = (\sum_{t,f} \left | (\mathbf{Y}^{\text{mag}}_{R}(t,f) - \mathbf{\hat{Y}}^{\text{mag}}_{R}(t,f))\right | ^2) / F,
\vspace{-0.1cm}
\end{equation}
where $\textbf{y}$ and $\hat{\textbf{y}}$ as the estimated and ground truth time domain signal, while $\textbf{Y}$ and $\hat{\textbf{Y}}$ denote their corresponding frequency domain spectrum. We conduct a fusion operation in Eq.~(\ref{outputfusion}) between estimated complex spectrum and magnitude since we find out that calculating the complex loss only using the estimated complex output without fusion may cause serious distortion, thus leading to extremely bad objective and subjective results.
\vspace{-0.5cm}
\section{Experiments and Results}
\vspace{-0.4cm}
\subsection{Datasets}
\vspace{-0.2cm}
In our experiments, the source speech data comes from LibriTTS~\cite{zen2019libritts}, AISHELL-3~\cite{shi2020aishell}, speech data of DNS challenge~\cite{reddy2021interspeech}, and the vocal part of MUSDB~\cite{rafii2017musdb18} while the source noise data comes from MUSAN~\cite{snyder2015musan}, noise data of DNS challenge, the music part of MUSDB, MS-SNSD~\cite{reddy2019scalable} and collected pure music data including classical and pop music. The training set contains 982 h source speech data and 230 h source noise data, while the development set contains 73 h source speech data and source 30 h noise data, respectively. 
Image method~\cite{allen1979image} is used to simulate RIR with RT60 ranges from 0.2 s to 1.2 s. 
Early reflection within 50 ms is used as the dereverberation training target. The training data are generated on-the-fly with 16k Hz sampling rate and segmented into 4 s chunks in one batch with SNR ranges from -5 to 15 dB. 
For model evaluation, three SNR ranges are selected to generate simulated dataset, namely [-5, 0], [0, 5] and [5, 10] dB, respectively. We simulate 500 noisy and reverberate pieces in each set. The source data among training, development and simulated evaluation set have no overlap. The blind test set of Interspeech 2021 DNS Challenge
%\footnote{https://github.com/microsoft/DNS-Challenge}
, which contains 600 noisy and reverberate pieces, is also selected as another evaluation dataset.

\vspace{-0.43cm}
\begin{table}[!h]
\centering
\caption{Results on different models in terms of PESQ, eSTOI, DNSMOS and MOS, where PESQ and eSTOI are calculated on simulated test set while DNSMOS and MOS are calculated on Interspeech2021 DNS challenge blind test set. Note that SDD-Net and DCCRN+ here are the results submitted to the challenge.}
\setlength{\tabcolsep}{2pt}
\label{results}
\scalebox{0.65}{
\begin{tabular}{lccccccccccccc}
\hline
\toprule
Model      & Cau. & \#Param. (M)                              &                      & PESQ                 &                                          &                      & eSTOI                &                      & DNSMOS  &MOS              \\
SNR (dB)    &  -    &    -                  
& {[}-5,0{]}           & {[}0,5{]}            & {[}5,10{]}           & {[}-5,0{]}           & {[}0,5{]}            & {[}5,10{]}           & -          &  -                                     \\ \midrule
Noisy      & -    & -                                       & 1.4710          & 1.7616          & 1.9904                   & 43.50          & 53.54          & 60.96          & 2.4139             & 1.8545                     \\
UFormer   & $\times$     & 9.46                                           & \textbf{2.4501} & \textbf{2.7472} & \textbf{2.9511} & \textbf{64.63} & \textbf{74.33} & \textbf{79.62} & \textbf{3.6032}     &   \textbf{3.3545}                 \\
UFormer   & $\checkmark$    &  9.46                                          & 2.4023          & 2.7265          & 2.9250                    & 64.22          & 74.29          & 79.46          & 3.5890      &   3.3523                       \\
~ \small -~FA & $\times$     & 9.02                                      & 2.4207          & 2.7273          & 2.9306                    & 64.19          & 74.11          & 79.37          &  3.5801       &-                               \\
~ \small -~DC & $\times$     & 9.31                                           & 2.3374          & 2.6689          & 2.8883                    & 62.86          & 73.03          & 78.52          &    3.5654     &-                                \\
~ \small \begin{tabular}[c]{@{}l@{}}-~encoder-decoder\\ ~ attention\end{tabular}  & $\times$     &  5.33                                          &   2.4218              &    2.7217             &    2.9177                        &   64.27             &    74.10            &  79.37              &   3.5381       &-                               \\
~ \small \begin{tabular}[c]{@{}l@{}}~~dilated conformer\\  \ \ $\rightarrow$ LSTM\end{tabular} &  $\times$    &  9.47                                          & 2.4106          & 2.7243          & 2.9258                   & 64.11          & 73.90          & 79.31          &    3.5839            &-                         \\
~ \small \begin{tabular}[c]{@{}l@{}}-~real-valued\\ ~ sub-modules\end{tabular} & $\times$    &  7.26                                          & 2.4266          & 2.7352          & 2.9402                   & 64.28          & 74.26          & 79.58          & 3.5751            &-                      \\
~ \small \begin{tabular}[c]{@{}l@{}}-~complex-valued\\ ~ sub-modules\end{tabular} & $\times$     & 3.85                                           & 2.4039          & 2.7025          & 2.9095                   & 63.55          & 73.37          & 78.78          & 3.5265          &-                      \\ \midrule
DCCRN~\cite{hu2020dccrn}      &  $\checkmark$    & 8.99                                           & 2.3652          & 2.6674          & 2.8676                    & 62.25          & 72.55          & 77.97          & 3.4915              &   3.2773                 \\
GCRN~\cite{tan2019learning}       & $\times$     & 30.83                                           & 2.2672          & 2.5768          & 2.7883                   & 61.43          & 71.87          & 77.70          &  3.3452          &-                            \\
PHASEN~\cite{yin2020phasen}     &  $\times$    & 8.41                                           & 2.3203          & 2.6170          & 2.8072                    & 62.76          & 72.73          & 78.12          &    3.4518               &-                      \\
SDD-Net~\cite{li2021simultaneous}    &  $\checkmark$    & 6.38    & -               & -               & -               & - & -                           & -              & \begin{tabular}[c]{@{}l@{}}3.36/3.47/\\ 3.56/\textbf{3.60}\end{tabular}  & 3.3432\\
DCCRN+~\cite{lv2021dccrn+}    &  $\checkmark$    & 4.71    & -               & -               & -               & -              & -              & -                            &  3.4260 & 3.0682\\
TasNet~\cite{luo2019conv}     & $\times$     &  8.69                                          & 2.2671          & 2.5649          & 2.7808                    & 61.11          & 71.30          & 77.50          &   3.3832              &-                        \\
DPRNN~\cite{luo2020dual}      & $\times$     &  2.60                                          & 2.2758          & 2.5723          & 2.7752                    & 61.30          & 71.61          & 77.10          &   3.2524        &-                              \\ 
\bottomrule \hline
\end{tabular}}
\vspace{-0.6cm}
\end{table}

\vspace{-0.4cm}
\subsection{Training Setups}
\vspace{-0.2cm}
For the input time domain signal, a 512-point short time Fourier transform (STFT) with a window size of 25 ms and a window shift of 10 ms is applied to each frame, resulting in 257 frequency bins. The channel numbers of encoder layers are [8, 16, 32, 64, 128, 128] and are inversed for decoder layers. The kernel size and stride for time and frequency axis are (2, 5) and (1, 2). The frame expansion in TA is 9 frames. For non-causal models, we combine the current frame with 4 historical and 4 future frames together as frame expansion. For causal models, we combine the current frame with 8 historical frames as frame expansion. For dilated conformer, the hidden dimension of two linear layers in FF are 64 and 128, respectively. The projection dimension and head number of both TA and FA are 16 and 1, respectively. For DC module, inspired by~\cite{li2021glance}, the output channel of the first Conv2d, D-Conv2d and the second Conv2d are 32, 32 and 128, respectively, to compress the feature dimension. The kernel size for time and frequency axis is (2, 1). The dilated conformer module is cascaded for 8 layers and the dilation expansion rate is 2. For causal models, dilated conformer only uses historical frames. For encoder-decoder attention, the kernel size of three Conv2ds is (2, 3). The dropout rate for all dropout layers in Uformer is 0.1. Both real and complex modules have the same configuration. Most recently, the effectiveness of power compressed spectrum has been demonstrated in the dereverberation task. We conduct the compression to magnitude before sending it into the network, and the compression variable is 0.5, which is the reported optimal parameter in ~\cite{li2021importance}. All models are trained with Adam optimizer for 15 epochs, thus all models are trained with 14730 h different noisy and reverberate speech data totally. The initial learning rate is 0.001 and will get halved if there is no loss decrease on the development set. Gradient clipping is applied with a maximum l2 norm of 5. $\alpha$, $\beta$, $\gamma$, $\zeta$ in hybrid loss are 5, 1/30, 1 and 1, respectively to make four losses are on the relatively same order of magnitude. 

\vspace{-0.45cm}
\subsection{Experiments}
\vspace{-0.19cm}
We conduct ablation experiments to prove the effectiveness of each proposed sub-modules, including a) Uformer without FA, b) Uformer without DC, c) Uformer without encoder decoder attention, d) substitute dilated complex \& real dual-path conformer with complex \& real LSTM, e) Uformer without all real-valued sub-modules (means we only model complex spectrum) and f) Uformer without all complex-valued sub-modules (means we only model magnitude). We also compare Uformer with DCCRN, PHASEN~\cite{yin2020phasen}, GCRN, SDD-Net~\cite{li2021simultaneous}, TasNet and DPRNN~\cite{luo2020dual}. DCCRN, PHASEN and GCRN are in complex domain, which aim to model magnitude and phase simultaneously. TasNet and DPRNN are advanced time domain speech front-end models. SDD-Net and DCCRN+ won the first and third prize of Interspeech 2021 DNS Challenge. The results of SDD-Net and DCCRN+ are the final results submitted to the challenge. All complex domain models are optimized with hybrid time and frequency domain loss while time domain models are optimized with SI-SNR loss. All the models are implemented with the best configurations mentioned in the corresponding literature.

\vspace{-0.4cm}
\subsection{Results and Analysis}
\vspace{-0.2cm}
Four evaluation metrics are used, namely perceptual evaluation of speech quality (PESQ), extended short-time objective intelligibility (eSTOI), DNSMOS~\cite{reddy2021dnsmos} and MOS. PESQ, and eSTOI are used to evaluate the objective performance of speech quality and intelligibility for simulated test set. DNSMOS is a non-intrusive perceptual objective metric, which is used to simulates the human subjective evaluation on DNS blind test set. We also organize 12 listeners to conduct a MOS test on 20 randomly selected clips from DNS blind test set to assess UFormer, DCCRN, DCCRN+ and SDD-Net.

The experimental results are shown in Table~\ref{results}. The four DNSMOS results of SDD-Net indicate the results of four training stages~\cite{li2021simultaneous}, namely denoising, dereverberation, spectral refinement and post-processing, respectively. Uformer gives the best performance in both objective and subjective evaluation, while the result of the causal version doesn't degrade generally. Compared with the time domain models, better performance is achieved for complex domain approaches, which indicates that it is more suitable to perform simultaneous enhancement and dereverberation in the T-F domain than the waveform level. Uformer reaches 3.6032 on DNSMOS, which is superior to all complex domain neural network based models and has relatively equal ability with the SDD-Net with post processing. However, compared with Uformer conducted in the form of end-to-end speech enhancement and dereverberation, the training procedure of SDD-Net is divided into four separate stages mentioned above, resulting in a relatively complicated training process. The idea of using dilated convolution layer shows great ability of modeling long range sequence dependencies, which lead to 0.08 and 0.04 average improvements on PESQ and DNSMOS, respectively. In addition, FA, encoder-decoder attention also contribute to a great extend.
Finally, the PESQ and DNSMOS degrade 0.02 and 0.04 when only modeling complex spectrum and degrade 0.03 and 0.08 when only modeling magnitude. 
These two results confirm the importance of the information interaction of complex spectrum and magnitude.
Uformer model also achieves the best MOS results (3.3545) among all SOTA complex domain models
%~\footnote{Demo page is available at https://felixfuyihui.github.io/Uformer-Demo/}.
\vspace{-0.45cm}
\section{Conclusion}
\vspace{-0.25cm}
In this paper, we propose Uformer~\footnote{Opensource code is available at https://github.com/felixfuyihui/Uformer} for simultaneous speech enhancement and dereverberation in both magnitude and complex domains. We leverage local and global contextual information as well as frequency information to improve the speech enhancement and dereverberation ability by dilated complex \& real dual-path conformer module. Hybrid encoder and decoder is adopted to model the complex spectrum and magnitude simultaneously. Encoder decoder attention is applied to better model the information interaction between encoder and decoder. Our experimental results indicate that the proposed UFormer reaches 3.6032 DNSMOS on the blind test set of Interspeech 2021 DNS Challenge, which surpasses the challenge best model -- SDD-Net.
%, which outperforms all SOTA time domain and complex domain speech front-end models on both objective and subjective performance.
% \section{Acknowledgement}
% The authors would like to thank A. Li from Institute of Acoustics, Chinese Academy of Sciences for providing the results of SDD-Net on DNS Challenge blind testset.
%\clearpage
\balance
\begin{spacing}{0.01}                    
\bibliographystyle{IEEE}
\bibliography{strings,refs}
\end{spacing}
\end{document}